\documentclass[11pt]{article}

\usepackage[margin=1in]{geometry}
\usepackage[T1]{fontenc}
\usepackage[utf8]{inputenc}
\usepackage{lmodern}
\usepackage{microtype}
\usepackage{amsmath,amssymb}
\usepackage{booktabs}
\usepackage{graphicx}
\usepackage{xcolor}
\usepackage{enumitem}
\usepackage{float}
\usepackage{placeins}
\usepackage[numbers,sort&compress]{natbib}
\usepackage[colorlinks=true,linkcolor=blue!55!black,citecolor=blue!55!black,urlcolor=blue!55!black]{hyperref}

\graphicspath{{../figures/}{./}}

\newcommand{\RegistryRootPackages}{2920}
\newcommand{\RegistryGraphNodes}{5503}
\newcommand{\RegistryDependencyEdges}{10491}
\newcommand{\RegistryCheckedEdges}{10006}
\newcommand{\RegistryRejectEdges}{1585}

\newcommand{\DownstreamTotalProbes}{2100}

\newcommand{\DownstreamRejectedClean}{1512}

\newcommand{\DownstreamRejectedConflicts}{629}
\newcommand{\DownstreamConflictPct}{41.60}

\newcommand{\DownstreamHeterogeneityP}{0.00}

\newcommand{\DriftLatestRejectPct}{15.84}
\newcommand{\DriftPreviousRejectPct}{46.52}
\newcommand{\DriftHistoricalRejectPct}{80.49}

\newcommand{\ClassifierPooledPpv}{66.67}
\newcommand{\ClassifierPooledNpv}{90.48}
\newcommand{\ErrorBudgetGraphBackwardMeanRejects}{20.24}
\newcommand{\ErrorBudgetGraphBackwardLowRejects}{15.00}
\newcommand{\ErrorBudgetGraphBackwardHighRejects}{26.00}
\newcommand{\ErrorBudgetAdoptBackwardAcceptLowPct}{55.56}
\newcommand{\ErrorBudgetAdoptBackwardAcceptHighPct}{74.60}

\newcommand{\AllenFiniteEcosystems}{4}

\newcommand{\ProtoConformanceRows}{300}
\newcommand{\ProtoConformanceMatches}{300}

\newcommand{\DirectionalPredictionRows}{29}
\newcommand{\TrajectoryAgeOnlyMae}{0.24}

\newcommand{\TrajectoryAgeOnlyGapMae}{0.42}

\newcommand{\TrajectoryInverseDirectionalMae}{0.13}

\newcommand{\TrajectoryInverseDirectionalGapMae}{0.07}

\newcommand{\TrajectoryResolverDirectionalMae}{0.18}

\newcommand{\DirectionalGapRows}{12}

\newcommand{\TrajectoryRows}{2995}
\newcommand{\TrajectoryLineages}{274}

\newcommand{\PermutationReplicates}{400}

\newcommand{\TemporalAgeOnlyBrier}{0.24}
\newcommand{\TemporalAgeOnlyAuc}{0.54}

\newcommand{\TemporalCheckerDirectionAuc}{0.57}

\newcommand{\TemporalResolverChannelBrier}{0.28}
\newcommand{\TemporalResolverChannelAuc}{0.51}

\newcommand{\TemporalFlipCandidates}{47}
\newcommand{\TemporalFlipEvents}{16}
\newcommand{\TemporalFlipCutoffDays}{365}
\newcommand{\TemporalFlipFinalDays}{1460}

\newcommand{\IntentionPypiAttempted}{249}
\newcommand{\IntentionPypiExcluded}{55}
\newcommand{\IntentionPypiLowerConflictPct}{70.28}
\newcommand{\IntentionPypiUpperConflictPct}{92.37}

\title{\vspace{-1.5em}\textbf{Interface-Variant Dynamics in Software Ecosystems:\\
Resolver-Induced Selection and Adoption in Package Graphs}}
\author{Faruk Alpay\thanks{Corresponding author: \texttt{alpay@lightcap.ai}} \quad Bar{\i}\c{s} Ba\c{s}aran\\[3pt]
\small Department of Computer Engineering, Bah\c{c}e\c{s}ehir University, Istanbul, T\"urkiye\\[-1pt]
\small \texttt{\{faruk.alpay, baris.basaran\}@bahcesehir.edu.tr}}
\date{}

\begin{document}
\maketitle

\begin{abstract}
\noindent
Compatibility research usually treats an interface change as a local
writer-reader decision.  Distributed software stacks make that decision
population structured: an RPC, telemetry, middleware, or service-contract
variant is introduced by one provider release and then spreads, stalls, or is
mediated across consumers, transitive dependencies, and resolver rules.  This
paper asks when that observation is a load-bearing software-engineering
estimator rather than evolutionary relabeling.  We mine interface histories,
audit npm, Maven Central, PyPI, and crates.io package graphs, execute
\DownstreamTotalProbes{} package-manager resolver probes, estimate an
ecosystem-specific selection coefficient $s$ from clean conflict
probabilities, and use that measured $s$ to forward evaluate a
pairwise-comparison absorbing process on the observed package graph.  We then
separate three evidential roles that are often conflated.  Fixation is a
forward evaluation, not independent evidence: once $s$ is measured, deviation
from $1/N$ follows mechanically from the non-neutral process.  Checker-derived
direction carries adoption signal: a direction-permutation null gives
checker-direction gap MAE 0.07 versus null median 0.43
($p=0.002$).  But because that direction is derived from the same boundary
state whose admitting frequency is predicted, it is a diagnostic rather than
an orthogonal selection test.  The stricter checker-free temporal test asks
whether early resolver-channel features predict later blocked-to-admitted
flips; in the present data they do not beat age-only
(Brier 0.28 versus 0.24, AUC 0.51 versus 0.54).  The result is therefore not a
closed claim that population dynamics have already been independently
validated.  It is a reproducible estimator audit that shows exactly where
resolver evidence becomes population input and where the current registry data
still fail to close the loop.
\end{abstract}

\section{Problem}

Generated interfaces evolve locally, but dependency managers expose their
effects at population scale in distributed software stacks.  A changed
protobuf, IDL, JSON protocol schema, or GraphQL schema creates a contract
variant on an RPC, middleware, telemetry, or service-client boundary.  Some
consumers remain on the old admissible variant, some accept the new one, and
some depend on resolver mediation at a version boundary.  The engineering
question is therefore not only whether one reader accepts one writer.  It is
whether a locally introduced provider variant can propagate through a
client-provider package graph under the update rule imposed by the ecosystem.

The population model is software-native.  Packages are carriers, contract
versions are variants, dependency neighborhoods define which distributed
clients, services, and middleware adapters can be affected, and
package-manager resolution supplies the local payoff of being across a
variant boundary.  The update event is adoption or imitation, not biological
reproduction.  That still puts the estimator in the mathematical family of
evolutionary graph theory and pairwise-comparison dynamics
\citep{lieberman2005evolutionary,ohtsuki2006simple,traulsen2006stochasticity,allen2017evolutionary}:
a variant starts rare, local interactions induce a payoff difference, and a
finite graph either absorbs at extinction or fixation.

The coupling is useful only if the estimator obeys an identifiability
discipline: a rule that measures selection must not contain an instance of the
selection signal it is asked to prove.  Clean resolver outcomes define $s$;
$s$ drives fixation simulations on the real graph as a forward evaluation; and
topology is reported as calibrated $\Delta\rho$ against graph nulls.  Those
steps put measured software costs onto a finite-population scale, but they do
not by themselves prove that adoption histories follow the same selection
channel.  The adoption analysis is therefore split into a checker-proximal
diagnostic, which tests whether the assignability direction has any real
adoption signal, and a checker-free temporal prediction, which tests whether
resolver-channel features predict later admission without reusing the checker
state as the answer.

The paper makes four concrete contributions:
\begin{enumerate}[leftmargin=1.5em]
  \item a calibrated directional-assignability instrument for generated
  interfaces, with accept-side gates rather than untested compatibility labels;
  \item an ecosystem-specific resolver selection estimator derived from real
  package-manager executions;
  \item fixation, topology, and power diagnostics on registry-scale package
  graphs, explicitly reported as forward consequences of measured $s$ rather
  than independent evidence; and
  \item a held-out adoption audit mined from time-stamped registry histories,
  including direction permutation, checker-free temporal flip prediction,
  hierarchical resolver/adoption consistency, and maintenance-recency
  stratification.
\end{enumerate}

\section{Interface Variants}

For a writer contract $W$ and reader contract $R$, write $W\preceq R$ when
every reader obligation is supplied by the writer with a compatible wire
representation.  The relation is directional: extra writer fields may be
ignored by an older reader, whereas missing reader fields and unsafe
wire-family changes reject the assignment.  The artifact evaluates exact
matching, the reader-obligation proxy, and FULL transitive compatibility.  The
proxy is not presented as a replacement for deployed Avro, protobuf,
Confluent, or DDS-XTypes semantics
\citep{avro2021spec,protobuf2026guide,confluent2026schema,omg2020xtypes};
it is a declared measurement instrument whose error is estimated before it is
used in population claims.

The source-history frame contains ROS 2 and OpenTelemetry interface histories,
plus a breakage-heavy schema-history corpus spanning gRPC, Apache Thrift,
Kafka protocol JSON, Confluent Schema Registry, and GraphQL histories.  These
are distributed communication surfaces rather than ordinary in-process APIs:
they define message, telemetry, broker, and service contracts that must remain
coherent across independently released clients and providers.  The histories
define the local variant classes and provide external compatibility checks
before the registry-scale graph audit.

\begin{table}[H]
  \centering
  \caption{Expanded breakage-heavy schema corpus.  These histories enlarge the
  empirical transition frame while keeping local proxy rules fixed.}
  \label{tab:expanded}
  \begin{tabular}{@{}llrrrr@{}}
\toprule
Corpus & Kind & Tags & Files & Changed & Breaking\\
\midrule
apache kafka protocol & kafka\_json & 58 & 281 & 232 & 97\\
apache thrift idl & thrift & 29 & 185 & 39 & 27\\
confluent schema registry & schema & 80 & 95 & 37 & 23\\
graphql js schema & graphql & 80 & 3 & 1 & 1\\
grpc proto & proto & 80 & 190 & 63 & 36\\
\bottomrule
\end{tabular}

\end{table}

\begin{figure}[H]
  \centering
  \includegraphics[width=0.78\linewidth]{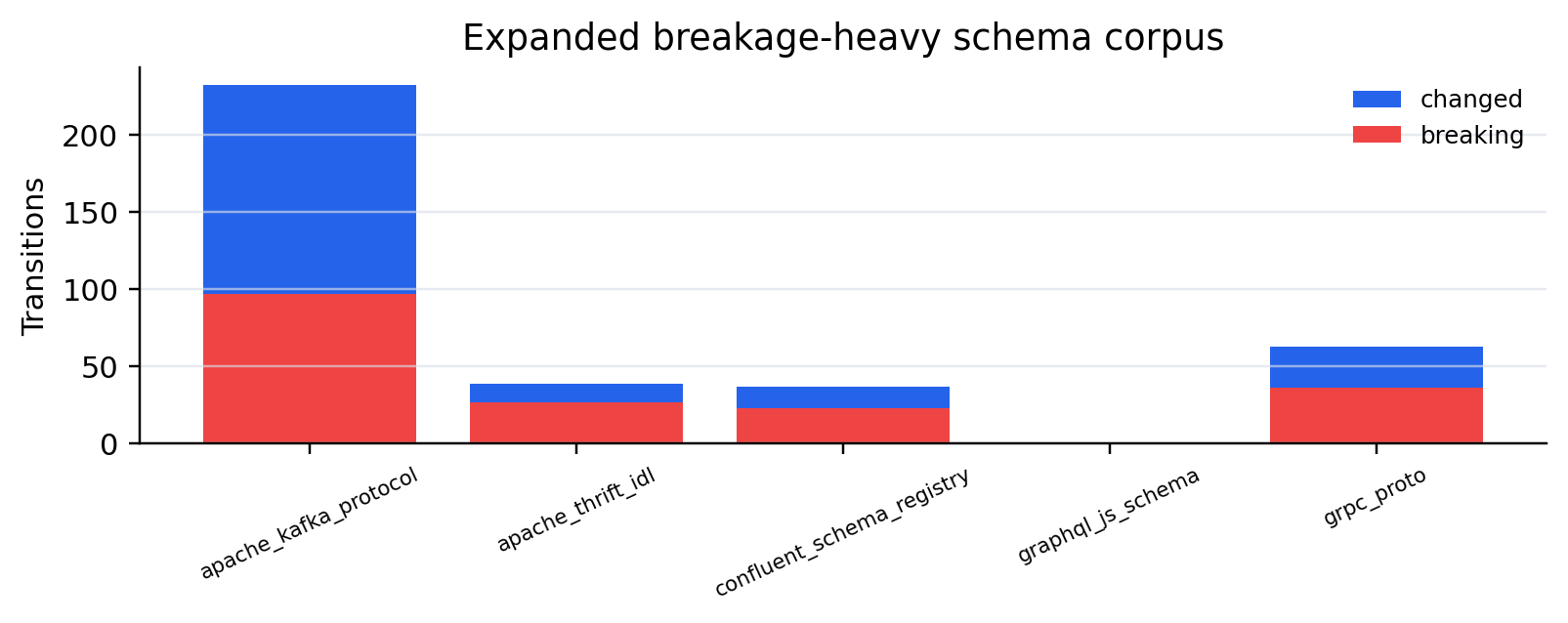}
  \caption{Changed and breaking transitions in the expanded schema corpus.}
  \label{fig:expanded}
\end{figure}

\section{Calibration Gates}

Population claims require the local compatibility instrument to be trustworthy
on both rejection and acceptance.  We therefore calibrate against three
independent references: observed OpenTelemetry protobuf transitions checked
with \texttt{buf breaking} at WIRE compatibility, synthetic Buf-WIRE
conformance cases generated from known-safe and known-breaking edit classes,
and ROS 2 IDL transitions evaluated with a DDS-XTypes-style backward
assignability checker.  The accept-side adoption analysis is enabled only when
negative predictive value clears a pre-stated gate.

\begin{table}[H]
  \centering
  \caption{Compatibility calibration gates.  Accept-side adoption quantities
  are reported only when accept NPV clears the gate.}
  \label{tab:calibration}
  \begin{tabular}{@{}lrrrr@{}}
\toprule
Calibration & $n$ & Reject PPV & Accept NPV & Gate\\
\midrule
Observed Buf-WIRE & 42 & 0.59 & 0.84 & pass\\
Buf-WIRE conformance & 300 & 1.00 & 1.00 & pass\\
XTypes independent states & 27 & 0.80 & 1.00 & pass\\
Empirical pooled & 69 & 0.67 & 0.90 & pass\\
\bottomrule
\end{tabular}

\end{table}

The calibration now supports the direction of the paper.  The empirical pooled
calibration gives a reject PPV of \ClassifierPooledPpv\% and an accept NPV of
\ClassifierPooledNpv\%, while the generated Buf-WIRE conformance suite checks
\ProtoConformanceRows{} cases with \ProtoConformanceMatches{} exact matches.
Classifier uncertainty is still propagated into graph and adoption quantities:
the backward graph-rejection count has bootstrap mean
\ErrorBudgetGraphBackwardMeanRejects{} with a
[\ErrorBudgetGraphBackwardLowRejects{}, \ErrorBudgetGraphBackwardHighRejects{}]
95\% interval, and the backward accept-rate interval is
\ErrorBudgetAdoptBackwardAcceptLowPct--\ErrorBudgetAdoptBackwardAcceptHighPct\%.

\section{Registry Graphs}

The registry audit asks whether a consumer's declared constraint still admits
the provider's current release.  That latest-provider rejection is a
variant-boundary position, not a build-failure label.  The audit covers npm,
Maven Central, PyPI, and crates.io roots and records both the package graph and
the checked constraint outcomes.

\begin{table}[H]
  \centering
  \caption{Registry-scale package graphs and latest-provider constraint checks.}
  \label{tab:registry}
  \begin{tabular}{@{}lrrrrr@{}}
\toprule
Ecosystem & Roots & Nodes & Edges & Checked & Latest rejects\\
\midrule
cargo & 700 & 935 & 2810 & 2810 & 257 (9.1\%)\\
maven & 750 & 1143 & 2879 & 2414 & 496 (20.5\%)\\
npm & 750 & 2100 & 2780 & 2780 & 578 (20.8\%)\\
pypi & 670 & 1351 & 2022 & 2002 & 254 (12.7\%)\\
all & 2920 & 5503 & 10491 & 10006 & 1585 (15.8\%)\\
\bottomrule
\end{tabular}

\end{table}

Across \RegistryRootPackages{} roots, the audit records
\RegistryGraphNodes{} graph nodes, \RegistryDependencyEdges{} dependency edges,
and \RegistryCheckedEdges{} checked constraints.  \RegistryRejectEdges{}
checked edges reject the provider's latest release.  The benign-drift baseline
separates this boundary position from ordinary version age: latest rejection is
\DriftLatestRejectPct\%, below the previous-release baseline
\DriftPreviousRejectPct\% and the historical-release baseline
\DriftHistoricalRejectPct\%.

\begin{table}[H]
  \centering
  \caption{Latest-provider rejection against benign release-history baselines.}
  \label{tab:drift}
  \begin{tabular}{@{}lrrrrr@{}}
\toprule
Ecosystem & Edges & Latest & Previous & Historical & Latest-prev.\\
\midrule
cargo & 2810 & 9.1\% & 23.1\% & 71.0\% & -13.9\\
maven & 2414 & 20.5\% & 94.7\% & 95.2\% & -4.8\\
npm & 2780 & 20.8\% & 66.0\% & 88.4\% & -45.0\\
pypi & 2002 & 12.7\% & 38.8\% & 78.9\% & -16.8\\
all & 10006 & 15.8\% & 46.5\% & 80.5\% & -25.8\\
\bottomrule
\end{tabular}

\end{table}

\section{Resolver Selection}

Local registry rejection becomes a population parameter only after executable
resolver behavior is measured.  Let $r^-_e$ be the probability that a clean
latest-reject probe in ecosystem $e$ still resolves successfully, and let
$r^+_e$ be the same probability for checker-accepted calibration probes.  With
Jeffreys smoothing,
\[
  \hat r^-_e = \frac{\mathrm{ok}^-_e + 1/2}{n^-_e + 1},
  \qquad
  \hat r^+_e = \frac{\mathrm{ok}^+_e + 1/2}{n^+_e + 1}.
\]
The resolver-derived selection coefficient is
\[
  s_e = \log \frac{\hat r^-_e}{\hat r^+_e}.
\]
Thus $s_e$ is not assigned by vocabulary.  It is the log-relative viability of
a boundary edge under the ecosystem's resolver.  Values near zero indicate
weak mediation cost; large negative values indicate that mixed-version
boundaries sharply reduce successful resolution relative to accepted
calibration edges.

\begin{table}[H]
  \centering
  \caption{Real resolver probes for latest-reject edges.  Clean excludes
  unavailable versions, registry outages, and timeouts.}
  \label{tab:downstream}
  \begin{tabular}{@{}lrrrrr@{}}
\toprule
Ecosystem & Attempted & Clean & Conflicts & Wilson 95\% & Excluded\\
\midrule
cargo & 246 & 246 & 2 (0.8\%) & [0.2, 2.9] & 0\\
maven & 496 & 496 & 376 (75.8\%) & [71.8, 79.4] & 0\\
npm & 578 & 576 & 76 (13.2\%) & [10.7, 16.2] & 2\\
pypi & 249 & 194 & 175 (90.2\%) & [85.2, 93.6] & 55\\
\bottomrule
\end{tabular}

\end{table}

Among \DownstreamRejectedClean{} clean latest-reject probes,
\DownstreamRejectedConflicts{} become resolver conflicts
(\DownstreamConflictPct\%).  The pooled rate is not averaged away because the
heterogeneity test is overwhelming ($p=\DownstreamHeterogeneityP$); the
heterogeneity is precisely the ecosystem-specific payoff regime that feeds
$s_e$.

Two sensitivity checks keep this measurement from becoming a clean-only
artifact.  First, the PyPI probe has \IntentionPypiExcluded{} excluded
latest-reject attempts out of \IntentionPypiAttempted{}; treating excluded
attempts as all successful or all conflicting gives an intention-to-probe
conflict range of \IntentionPypiLowerConflictPct--\IntentionPypiUpperConflictPct\%,
so the extreme PyPI penalty is not created by dropping unavailable versions.
Second, the pooled reject PPV is \ClassifierPooledPpv\%, so the selection
coefficient is recomputed under a conservative mixture in which false rejects
resolve like accepted calibration edges.  This propagates reject-side
classifier error into $s_e$ itself rather than only into edge counts.

\begin{table}[H]
  \centering
  \caption{Resolver-selection sensitivity.  ITT bounds treat unclean
  latest-reject probes as all non-conflicts or all conflicts; PPV-adjusted
  $s$ subtracts the pooled false-reject component from the boundary stratum.}
  \label{tab:selection-sensitivity}
  \resizebox{\linewidth}{!}{\begin{tabular}{@{}lrrrr@{}}
\toprule
Ecosystem & Attempted & Clean conflict & ITT conflict bounds & PPV-adjusted $s$\\
\midrule
npm & 578 & 13.2\% & 13.1--13.5\% & -0.20\\
maven & 496 & 75.8\% & 75.8--75.8\% & -6.77\\
pypi & 249 & 90.2\% & 70.3--92.4\% & -5.85\\
cargo & 246 & 0.8\% & 0.8--0.8\% & -0.01\\
\bottomrule
\end{tabular}
}
\end{table}

\section{Fixation Model}

For each ecosystem, the directed dependency graph is projected to the largest
undirected interaction component $G_e=(V_e,E_e)$.  Each package has state
$x_i\in\{0,1\}$, where 1 is the introduced variant.  Let
$m_i(x)$ be the fraction of neighbors of $i$ with the opposite state.  The
software payoff is
\[
  \pi_i(x) = s_e\,m_i(x).
\]
At each update, a focal package $i$ samples a neighboring model package $j$ and
copies $j$ with Fermi probability
\[
  P(i\leftarrow j) =
  \frac{1}{1+\exp[-(\pi_j(x)-\pi_i(x))]}.
\]
The process starts with one introduced variant and stops at extinction or
fixation.  The measured output is $\rho(s_e,G_e)$, estimated by Monte Carlo
with Wilson intervals.  The neutral benchmark is exact: when $s_e=0$, the
single-variant fixation probability is $1/|V_e|$.

\begin{table}[H]
  \centering
  \caption{Resolver-derived selection and graph fixation.  The same measured
  $s$ from resolver probes is used in the absorbing process.}
  \label{tab:selection-fixation}
  \resizebox{\linewidth}{!}{\begin{tabular}{@{}lrrrl@{}}
\toprule
Ecosystem & Resolver $s$ [95\%] & Forward $\rho(s,G)$ [95\%] & Neutral $1/N$ & Regime\\
\midrule
npm & -0.131 [-0.169, -0.094] & 0.00017 [0.00007, 0.00039] & 0.00059 & suppression\\
maven & -1.286 [-1.457, -1.118] & 0.00000 [0.00000, 0.00013] & 0.00108 & suppression\\
pypi & -2.185 [-2.670, -1.808] & 0.00000 [0.00000, 0.00013] & 0.00106 & suppression\\
cargo & -0.007 [-0.024, 0.008] & 0.00063 [0.00041, 0.00099] & 0.00114 & near-neutral\\
\bottomrule
\end{tabular}
}
\end{table}

\begin{figure}[H]
  \centering
  \includegraphics[width=\linewidth]{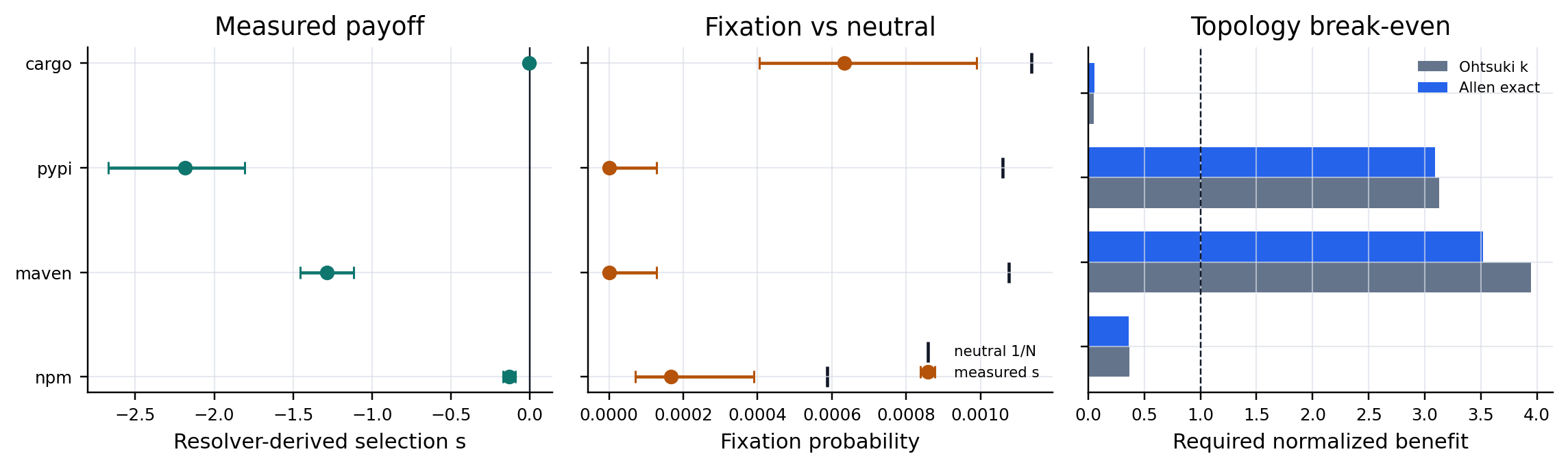}
  \caption{Measured resolver selection, fixation against the neutral $1/N$
  benchmark, and topology break-even requirements.}
  \label{fig:resolver-evolution}
\end{figure}

This table is a forward evaluation, not independent evidence for the
population hypothesis.  Once $s_e$ is measured, separation from $1/N$ is the
expected consequence of running a non-neutral absorbing process.  The result is
still useful because it puts the resolver penalty on the registry graph scale,
but it is not counted as a second empirical test of the same signal.  The
cargo row shows why this distinction matters: its resolver interval crosses
zero, so the correct interpretation is near-neutral even though the point
Monte Carlo interval falls below $1/N$.  Section~\ref{sec:adoption-validation}
therefore treats later registry histories as a separate audit and distinguishes
checker-proximal adoption signal from checker-free prediction.

\section{Topology Conditions}

Topology is reported as a calibrated response rather than a pass/fail
population claim.  First, the observed graph is evaluated under Ohtsuki's
degree-based rule and Allen et al.'s structure-coefficient formalism.  These
are diagnostic thresholds for a benefit-favoring regime; the measured resolver
regime here is non-positive in all four ecosystems, so the operational target
is suppression or near-neutrality, not cooperation amplification.  Let $c_e$
be the incremental conflict cost measured by the resolver audit.  The Ohtsuki
diagnostic asks whether the required benefit exceeds $c_e\bar{k}$.  For the
Allen diagnostic, coalescence samples estimate the structure threshold
\[
  (b/c)^* = \frac{t_2}{t_3-t_1},
\]
with infinite threshold when the denominator is nonpositive; the required
benefit is $c_e(b/c)^*$.  Second, fixation response on the observed graph is
compared with three null families: degree-preserving rewiring,
clustering-preserving rewiring, and a directed consumer-provider configuration
null projected back to an interaction graph.  The reported quantity is
$\Delta\rho$, the observed signed fixation response minus the median null
response.

\begin{table}[H]
  \centering
  \caption{Topology diagnostics and calibrated null response.  Positive
  $\Delta\rho$ means the observed graph strengthens the measured selection
  response relative to the null median; negative values mean suppression.}
  \label{tab:structure}
  \resizebox{\linewidth}{!}{\begin{tabular}{@{}lrrrrrr@{}}
\toprule
Ecosystem & $N$ & $\bar{k}$ & Allen $(b/c)^*$ & $\Delta\rho_{deg}$ & $\Delta\rho_{clust}$ & $\Delta\rho_{c/p}$\\
\midrule
npm & 1704 & 2.98 & 2.90 & -0.00017 & -0.00017 & -0.00017\\
maven & 929 & 6.17 & 5.50 & 0.00000 & 0.00000 & 0.00000\\
pypi & 943 & 3.93 & 3.88 & 0.00000 & 0.00000 & 0.00000\\
cargo & 879 & 5.88 & 6.88 & 0.00003 & 0.00020 & 0.00070\\
\bottomrule
\end{tabular}
}
\end{table}

\AllenFiniteEcosystems{} ecosystems have finite Allen thresholds under the
estimated coalescence coefficients, but the measured data do not enter the
positive-benefit branch those thresholds favor.  The null comparisons instead
show whether the observed package graph changes the suppression response
relative to matched ensembles.  This makes topology load-bearing only where it
has an operational target: the sign and magnitude of $\Delta\rho$ under the
measured resolver regime.

\section{Adoption Validation}
\label{sec:adoption-validation}

The fixation model is a finite-population endpoint.  The adoption histories
are the only place where the population claim can be checked against later
registry behavior rather than replayed from the measured resolver coefficient.
We mine \TrajectoryRows{} consumer-release observations over
\TrajectoryLineages{} provider-boundary lineages.  A lineage has
\[
  a_\ell =
  \begin{cases}
    +1, & \text{provider-favored: the provider boundary is admitted},\\
    -1, & \text{consumer-favored: the consumer constraint blocks it}.
  \end{cases}
\]
This label is generated by the checker state of the edge.  That makes it a
useful engineering quantity, but it also creates the central identifiability
risk: if the outcome is admitting frequency, then a predictor that contains
the same admission state is not independent evidence for selection.  The
analysis below therefore separates a checker-proximal diagnostic from a
checker-free temporal prediction.

First, we ask whether checker direction carries any real adoption signal after
holding out ecosystems.  The baseline is
\[
  \operatorname{logit}\hat{x}_{\ell e}(t)
  = \alpha + \beta \log(1+t/90).
\]
The resolver-directional edge game adds the measured population penalty
$|s^{\mathrm{pop}}_e|=|q_e s_e|$ with its sign supplied by assignability:
\[
  \operatorname{logit}\hat{x}_{\ell e}(t)
  = \alpha + \beta \log(1+t/90)
  + \gamma a_\ell |s^{\mathrm{pop}}_e|.
\]
The checker-direction model estimates a free direction effect,
\[
  \operatorname{logit}\hat{x}_{\ell e}(t)
  = \alpha + \beta \log(1+t/90) + \eta a_\ell,
\]
and is reported as a diagnostic, not as an independent inverse estimator,
because $a_\ell$ is derived from the checker state.

\begin{table}[H]
  \centering
  \caption{Leave-one-ecosystem-out adoption diagnostics.  The direction-gap
  MAE is computed on the \DirectionalGapRows{} ecosystem-horizon cells where
  both provider- and consumer-favored lineages are observed.}
  \label{tab:trajectory}
  \begin{tabular}{@{}lrrrr@{}}
\toprule
Held-out model & Rows & MAE & Final MAE & Direction-gap MAE\\
\midrule
Age-only & 29 & 0.24 & 0.21 & 0.42\\
Resolver-directional & 29 & 0.18 & 0.19 & 0.39\\
Checker-direction & 29 & 0.13 & 0.16 & 0.07\\
\bottomrule
\end{tabular}

\end{table}

\begin{figure}[H]
  \centering
  \includegraphics[width=0.92\linewidth]{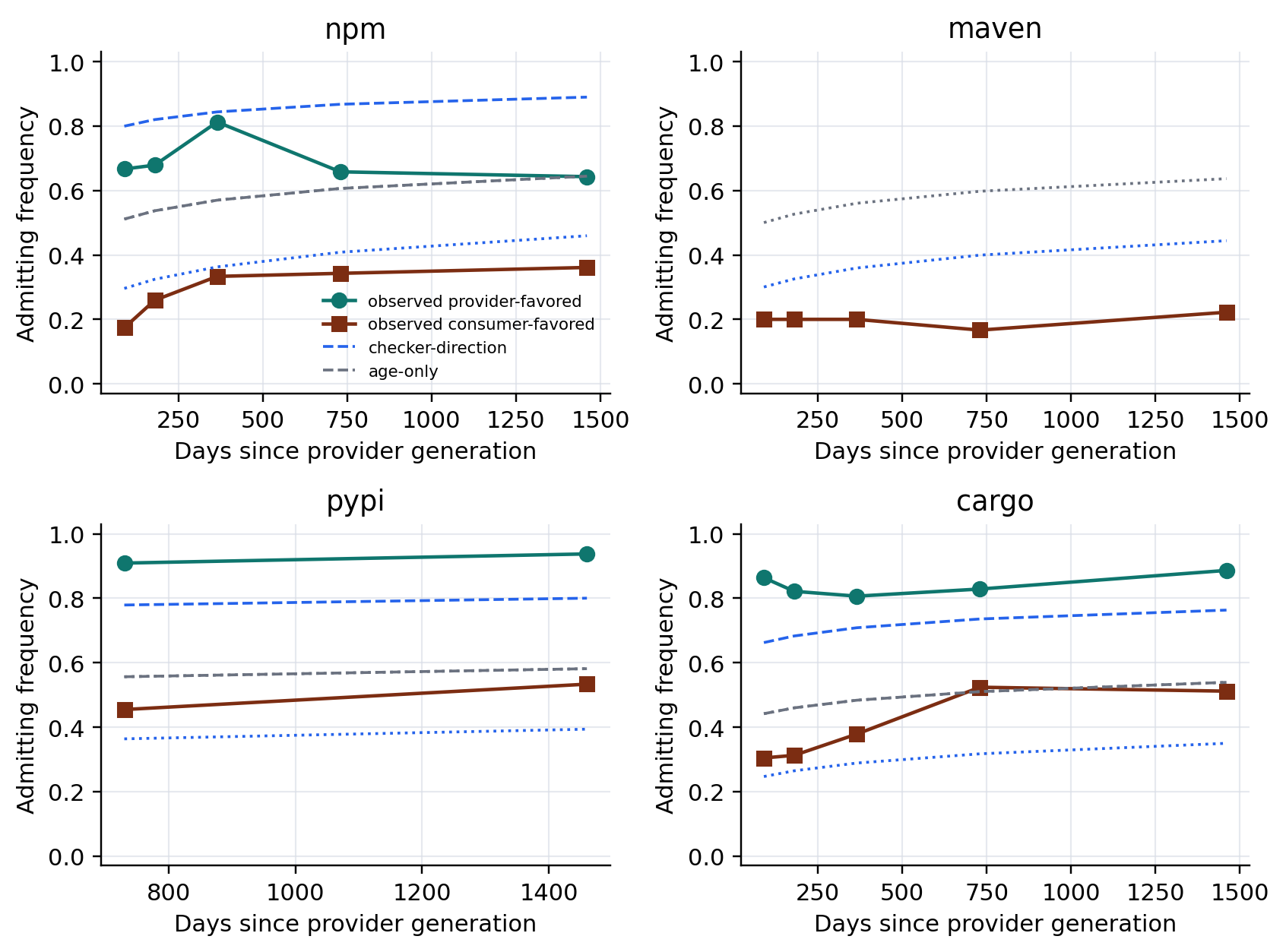}
  \caption{Observed adoption trajectories and leave-one-ecosystem-out
  predictions.  Solid lines are observed provider- and consumer-favored
  lineages; dashed and dotted lines are checker-direction and age-only
  predictions.}
  \label{fig:trajectory}
\end{figure}

The held-out ecosystem fits none of its own coefficients.  Across
\DirectionalPredictionRows{} held-out directional rows, age-only MAE is
\TrajectoryAgeOnlyMae{}, the resolver-directional edge game improves it to
\TrajectoryResolverDirectionalMae{}, and the checker-direction diagnostic
improves it to \TrajectoryInverseDirectionalMae{}.  The equal-age
direction-gap error falls from \TrajectoryAgeOnlyGapMae{} to
\TrajectoryInverseDirectionalGapMae{}.  That is real signal in the checker
state, but it is not yet an orthogonal population estimate: age-only predicts
zero direction gap by construction, and the direction label is derived from
the same edge state whose admitting frequency is being summarized.

To quantify this limitation, we run \PermutationReplicates{} direction
permutations that swap provider- and consumer-favored labels within
ecosystem-horizon cells while preserving the observed adoption frequencies and
sample sizes.  The checker-direction row remains far outside this null, while
the resolver-directional row does not.

\begin{table}[H]
  \centering
  \caption{Direction-permutation diagnostic.  Low observed gap MAE means the
  model explains the provider-minus-consumer adoption gap.  The null swaps
  direction labels within ecosystem-horizon cells.}
  \label{tab:direction-permutation}
  \begin{tabular}{@{}lrrrr@{}}
\toprule
Model & Observed gap MAE & Null median & Null 95\% & $p_{perm}$\\
\midrule
Age-only & 0.42 & 0.42 & [0.42, 0.42] & 1.000\\
Resolver-directional & 0.39 & 0.42 & [0.35, 0.52] & 0.177\\
Checker-direction & 0.07 & 0.43 & [0.25, 0.52] & 0.002\\
\bottomrule
\end{tabular}

\end{table}

This table rules out the trivial explanation that any arbitrary direction
label would fit the adoption gap.  It does not rule out signal leakage from
the checker into the outcome.  The stricter test removes checker direction
from the predictor.  At \TemporalFlipCutoffDays{} days after a provider
generation, we retain lineages that are still not admitted and ask whether
they flip to admitted by \TemporalFlipFinalDays{} days.  The resolver-channel
model uses only age, the observed resolver outcome, and the ecosystem-level
resolver penalty; it does not use $a_\ell$.

\begin{table}[H]
  \centering
  \caption{Checker-free temporal flip prediction.  Candidates are lineages
  blocked at the cutoff; flips are later blocked-to-admitted transitions by
  the final horizon.  Lower Brier and higher AUC are better.}
  \label{tab:temporal-flip}
  \begin{tabular}{@{}lrrrrr@{}}
\toprule
Model & Candidates & Flips & Brier & AUC & Top-half flip\\
\midrule
Age-only & 47 & 16 & 0.24 & 0.54 & 0.35\\
Resolver-channel & 47 & 16 & 0.28 & 0.51 & 0.26\\
Checker-direction & 47 & 16 & 0.24 & 0.57 & 0.39\\
\bottomrule
\end{tabular}

\end{table}

This is the strictest adoption check in the present artifact.  On
\TemporalFlipCandidates{} candidate lineages with \TemporalFlipEvents{}
later flips, the resolver-channel model does not beat age-only: its Brier
score is \TemporalResolverChannelBrier{} versus \TemporalAgeOnlyBrier{}, and
its AUC is \TemporalResolverChannelAuc{} versus \TemporalAgeOnlyAuc{}.  The
checker-direction diagnostic remains slightly better ranked
(\TemporalCheckerDirectionAuc{} AUC), but that is exactly the
checker-proximal signal separated above.  Therefore the registry adoption
data do not yet license the stronger claim that resolver-measured selection
independently predicts future admission.

\begin{table}[H]
  \centering
  \caption{Checker-direction selection from adoption histories.  The
  provider/consumer column gives final-horizon lineage counts; $|s_{resolver}|$
  is the boundary resolver penalty and $|s_{pop}|$ multiplies it by boundary
  exposure.}
  \label{tab:inverse-selection}
  \resizebox{\linewidth}{!}{\begin{tabular}{@{}lrrrr@{}}
\toprule
Ecosystem & Provider/consumer lineages & $\hat{s}_{check}$ [95\%] & $|s_{resolver}|$ & $|s_{pop}|$\\
\midrule
npm & 42/36 & 0.81 [0.60, 1.05] & 0.131 & 0.027\\
maven & 3/18 & 1.48 [0.93, 7.57] & 1.286 & 0.264\\
pypi & 16/15 & 1.38 [0.78, 7.82] & 2.185 & 0.277\\
cargo & 44/43 & 1.01 [0.78, 1.29] & 0.007 & 0.001\\
\bottomrule
\end{tabular}
}
\end{table}

The ecosystem-level checker-direction estimates are positive in all four
ecosystems, but a four-point Pearson correlation is not an inferential result.
We therefore replace the raw correlation claim with a small hierarchical
measurement model that treats both the adoption-side estimate and the
resolver-side penalty as noisy observations.  The coupling slope is positive
in the posterior median, while residual heterogeneity remains large enough
that the result should be read as consistency, not closure.

\begin{table}[H]
  \centering
  \caption{Hierarchical resolver/adoption consistency.  The model regresses
  checker-direction adoption estimates on resolver boundary penalties while
  propagating both measurement intervals and residual heterogeneity.}
  \label{tab:hierarchical-consistency}
  \begin{tabular}{@{}lrrr@{}}
\toprule
Quantity & Median & 95\% interval & Posterior probability\\
\midrule
Coupling slope $\beta$ & 0.42 & [0.02, 1.30] & $P(\beta>0)=0.98$\\
Residual heterogeneity $\tau$ & 0.29 & [0.00, 1.34] & $P(\tau<0.5)=0.70$\\
\bottomrule
\end{tabular}

\end{table}

The mismatch is not an error to be hidden.  Cargo has near-zero resolver
penalty but a strong checker-direction adoption gap, while Maven has only
three provider-favored lineages and therefore a wide adoption interval.  A
maintenance-recency stratification shows another confound: stale providers can
have very different admitting frequencies from actively maintained providers,
so adoption pressure is not reducible to resolver mechanics alone.

\begin{table}[H]
  \centering
  \caption{Maintenance-recency stratification for final lineage states.
  Active providers have a latest release within one year of the observation
  date; stale providers do not.}
  \label{tab:maintenance-recency}
  \resizebox{\linewidth}{!}{\begin{tabular}{@{}llrrr@{}}
\toprule
Ecosystem & Provider recency & Lineages & Provider-favored & Admit freq.\\
\midrule
cargo & active provider & 72 & 30 & 0.64\\
cargo & stale provider & 20 & 18 & 0.90\\
maven & active provider & 19 & 3 & 0.63\\
maven & stale provider & 29 & 1 & 0.45\\
npm & active provider & 66 & 32 & 0.44\\
npm & stale provider & 19 & 13 & 0.79\\
pypi & active provider & 41 & 17 & 0.80\\
pypi & stale provider & 8 & 8 & 1.00\\
\bottomrule
\end{tabular}
}
\end{table}

\section{Power Diagnostic}

The fixation power calculation is a design diagnostic, not an empirical
validation.  It prevents small toy graphs from being mistaken for population
evidence by computing the minimum detectable positive and negative mixed-edge
selection strength for an exact well-mixed pairwise-comparison fixation
probability, evaluated at canonical graph sizes and at the actual registry
component sizes.  The canonical calculation falls from roughly
$|s|\approx 0.043$ at $N=8$ to $|s|\approx 0.013$ at $N=64$ and
$|s|\approx 0.0036$ around $N=1024$, which is why tiny release-history graphs
are treated as descriptive only.  At the observed registry component sizes
($N=879$ to $1704$), the 80\% minimum detectable selection strengths are
approximately $0.0020$--$0.0030$ for positive selection and
$-0.0029$--$-0.0038$ for negative selection.

\begin{table}[H]
  \centering
  \caption{Minimum detectable selection strength for the registry-scale
  fixation test at 80\% power.}
  \label{tab:power}
  \begin{tabular}{@{}lrrrr@{}}
\toprule
Ecosystem & $N$ & Trajectories & MDE $s>0$ & MDE $s<0$\\
\midrule
cargo & 879 & 30000 & 0.0030 & -0.0038\\
maven & 929 & 30000 & 0.0029 & -0.0036\\
pypi & 943 & 30000 & 0.0028 & -0.0038\\
npm & 1704 & 30000 & 0.0020 & -0.0029\\
\bottomrule
\end{tabular}

\end{table}

\section{Threats To Validity}

The registry audit observes declared dependency constraints and resolver
plans, not downstream source builds.  Optional extras, native build scripts,
environment-specific test suites, and undeclared runtime coupling remain
outside this cross-registry design.  Ecosystem-specific mechanisms are only
partly represented: npm peer dependencies and overrides, Maven
\texttt{dependencyManagement} and BOM import scope, PyPI environment markers
and extras, and Cargo feature unification can all alter the realized boundary
state observed by a downstream build.

The directed dependency graph must be projected to an interaction graph for
pairwise comparison; the consumer-provider configuration null is included to
test the projection's most obvious structural loss.  The endpoint fixation
model still uses a symmetric mixed-boundary payoff and is therefore a forward
diagnostic, not the decisive population test.  The adoption validation relaxes
that symmetry by carrying assignability direction into the edge game, but that
direction is checker-derived.  Consequently the checker-direction results are
diagnostic of adoption signal, while the checker-free temporal flip test is
the stricter estimator check and is negative in this snapshot.

Calibration corpora are larger than the original release-history frame but
still partial, so classifier error is propagated rather than ignored.  The
registry snapshot is frozen at the observation date, and the artifact records
the installed tool versions used for resolver probes.  This makes the
reported run reproducible, but it does not remove registry drift as a threat
to future reruns.

\section{Artifact}

The arXiv source package places the article source and root figures at the
submission root and records code, data, and reproducibility notes under
\texttt{anc/}.  The ancillary directory contains the mined histories, registry
graphs, resolver probes, calibration outputs, fixation runs, topology
classifications, trajectory predictions, permutation nulls, temporal flip
predictions, hierarchical consistency summaries, maintenance-recency
strata, tool-version manifests, and table-regeneration code.  Table generation
is offline from recorded CSV/JSON outputs; full reproduction reruns mining
against upstream Git repositories and package registries.

\section{Related Work}

Software evolution work established that useful systems keep changing and that
API migration imposes developer cost
\citep{lehman1980programs,parnas1994software,dig2006api,bogart2016break,kula2018developers}.
Semantic versioning communicates intent but does not implement assignability
or resolver mediation \citep{prestonwerner2013semver}.  Avro, Protocol
Buffers, Confluent Schema Registry, and DDS-XTypes provide local compatibility
baselines
\citep{avro2021spec,protobuf2026guide,confluent2026schema,omg2020xtypes}.
ROS 2 and DDS provide the distributed middleware setting for generated
interfaces and typed message exchange
\citep{quigley2009ros,macenski2022ros2,omg2015dds}.  Evolutionary graph theory
and pairwise-comparison dynamics provide the finite-population language for
variants, structured interaction, selection, fixation, and drift
\citep{lieberman2005evolutionary,ohtsuki2006simple,traulsen2006stochasticity,allen2017evolutionary}.
The contribution here is to instantiate that language with package-manager
semantics and executable resolver evidence, while exposing where the resulting
estimator is and is not identifiable from registry histories.

\section{Conclusion}

Interface evolution in package ecosystems can be analyzed as
contract-variant dynamics, but only when the software measurements are allowed
to constrain the population model rather than merely rename its terms.  This
paper makes that boundary explicit.  Resolver conflicts estimate
ecosystem-specific selection, the real package graph defines a forward
fixation diagnostic, and topology is reported as calibrated $\Delta\rho$
against graph nulls.  Adoption histories show that checker-derived direction
contains strong adoption signal, and a permutation null confirms that the
signal is not an arbitrary label effect.  The stricter checker-free temporal
test, however, does not yet show that resolver-channel features independently
predict future admission.  The artifact therefore turns relabeling into a
falsifiable estimator pipeline, but it also records the present failure point:
the resolver-to-adoption loop is measured, diagnosable, and not yet closed by
this snapshot.

\bibliographystyle{plainnat}
\bibliography{refs}

\end{document}